%% file: elsarticle-template-num.tex
\documentclass[authoryear,preprint,12pt,a4paper]{elsarticle}

\usepackage{amssymb}

\usepackage[utf8]{inputenc}
\usepackage{algorithm}
\usepackage{algpseudocode}
\usepackage{amsmath,amssymb,xcolor,soul,subcaption}
\usepackage{hyperref}
\usepackage[normalem]{ulem}
\usepackage[margin=2.5cm]{geometry}
\usepackage{makecell}
\usepackage{mathtools}
\usepackage{booktabs}
\usepackage[T1]{fontenc}
\usepackage{soul}
\usepackage{rotating} 
\usepackage{amsmath}
\usepackage{bbm}

\useunder{\uline}{\ul}{}
\usepackage{lscape}
\usepackage{tikz}
\def\checkmark{\tikz\fill[scale=0.4](0,.35) -- (.25,0) -- (1,.7) -- (.25,.15) -- cycle;}

\newcommand{\mycomment}[1]{}

\DeclareMathOperator{\EX}{\mathbb{E}}

\journal{Journal}

\begin{document}

\begin{frontmatter}



\title{A review on reinforcement learning methods for mobility on demand systems}


\author[irtx,cs]{Tarek Chouaki}
\author[irtx]{Sebastian Hörl}
\author[em,cs]{Jakob Puchinger}

\address[irtx]{Institut de Recherche Technologique SystemX, 91220, Palaiseau, France}

\address[cs]{Université Paris-Saclay, CentraleSupélec, Laboratoire Génie Industriel, 91190, Gif-sur-Yvette, France}

\address[em]{EM Normandie Business School, Métis Lab, 92110, Clichy, France}

\begin{abstract}

\input{0_Abstract}
\end{abstract}




\end{frontmatter}

\input{1_Introduction}
\input{2_Background}

\input{3_Literature}
\input{4_Analysis}

\input{Acknowledgements}

\bibliographystyle{elsarticle-harv} 
\bibliography{main.bib}




\end{document}

%% file: 0_Abstract.tex
Mobility on Demand (MoD) refers to mobility systems that operate on the basis of immediate travel demand. Typically, such a system consists of a fleet of vehicles that can be booked by customers when needed. The operation of these services consists of two main tasks: deciding how vehicles are assigned to requests (vehicle assignment); and deciding where vehicles move (including charging stations) when they are not serving a request (rebalancing). A field of research is emerging around the design of operation strategies for MoD services, and an increasingly popular trend is the use of learning based (most often Reinforcement Learning) approaches. We review, in this work, the literature on algorithms for operation strategies of MoD systems that use approaches based on Reinforcement Learning with a focus on the types of algorithms being used. The novelty of our review stands in three aspects: First, the algorithmic details are discussed and the approaches classified in a unified framework for sequential decision-making. Second, the use cases on which approaches are tested and their features are taken into account. Finally, validation methods that can be found across the literature are discussed. The review aims at advancing the state of the art by identifying similarities and differences between approaches and highlighting current research directions. 

%% file: 1_Introduction.tex
\section{Introduction}
Mobility on Demand (MoD) refers to mobility systems that operate on the basis of travelers' immediate demand,  
in contrast to regular public transit systems that follow fixed routes and schedules. 
They consist of a fleet of vehicles that can be booked by customers when needed to perform trips,
and the operation of such services is composed of two main tasks: deciding how vehicles are assigned to requests (vehicle assignment); and deciding where vehicles move (including charging stations) when they are not serving a request (rebalancing).



The research around operation strategies for MoD strongly intersects with operations research, and many well established approaches from that domain have been adapted and tested for MoD. \citet{mourad2019survey} systematizes the algorithmic details of various optimization-based operational strategies related to MoD by considering the problems that are addressed, the constraints that are taken into account (capacity, time cost...), the objective functions of the algorithms and their complexities as well as the concerned use cases. \citet{zardini2022analysis} gives a similar overview without detailing the objective of each approach. 

On the other hand, heuristic-based operation strategies have been proposed \citep{ruch2018amodeus, de2020optimization, alonso-mora_-demand_2017, hyland2018dynamic, horl_fleet_2019}. They are frequently applied in studies using agent-based simulations of MoD systems that examine the impact of MoD operation strategies on the transport system performance (customer waiting times, empty vehicle distance). A major focus of research in those studies is the fleet size required to achieve a certain performance level under a well-defined demand level \citep{ruch2018amodeus}.

The development of MoD is closely related to the evolution of Autonomous Vehicle (AV) technology. \citet{jing_agent-based_2020} present a systematic literature review on agent-based simulations of AVs. The review assesses which  operational aspects such as recharging, relocation and ride-sharing are taken into account in the literature, but does not detail in depth the differences between the algorithms and their implementation. Similarly, \citet{chakraborty2021review} reviews relevant vehicle-assignment algorithms with a focus on their impact on different stakeholders.

An emerging sub-field of research on MoD fleet control is the use of Reinforcement Learning (RL) based approaches. 
Instead of manually implementing strategies for the MoD system to follow, these approaches allow the MoD fleet to learn strategies and how to take decisions that maximize the vehicle's performance, transfer learned strategies to new use cases, allow flexibility in the objective functions and take into account  any partially available information depending on the use case. The present review intends to synthesize the reinforcement learning approaches and use cases that have been studied to find pathways for future research. 

%% file: 2_Background.tex
\section{Background}
\label{background}
In this section, we first focus on Reinforcement Learning (RL) and present the different types of approaches that we find in the literature of MoD. We then present a recent framework under which methods for sequential decision-making under uncertainty can be classified and compared.  


\subsection{Reinforcement Learning}
\label{reinforcement_learning}
The term reinforcement learning refers to three separate but related aspects: a class of problems, the algorithms for solving those problems, and the field of research that studies the design of algorithms to solve these problems\citep{sutton_reinforcement_1998}.

A RL problem is any problem where we consider an \textit{agent} or multiple agents evolving in an \textit{environment} and interacting with it through
a loop of \textit{actions} performed by the agent(s) and a \textit{reward} signal returned by the environment indicating how well the agent has performed. More formally, RL problems can be represented with Markov Decision Processes (MDP)\citep{littman_markov_1994}  which consist of the following components: (i) a set of states $S$, (ii) a set of actions $A$, (iii) a transition function $p(s'|s,a) = Pr(S_{t+1}=s'|S_t=s,A_t=a)$ where $S_{t+1}$, $S_t$ and $A_t$ respectively indicate the state at step $t+1$, the state at step $t$ and the action taken at step $t$ after observing $S_t$ and before observing $S_{t+1}$, and (iv) a reward function $r(s,a,s') = \EX(R_{t+1}|S_t=s,A_t=a,S_{t+1}=s')$ where $R_{t+1}$ indicates the reward obtained at time $t+1$ after observing $S_t$ and performing $A_t$ and arriving at $S_{t+1}$. 

The agent then needs to learn, through trial and error, a \textit{policy} $P(s,a)=Pr(A_{t}=a|S_{t}=s)$ specifying which action to take given the \textit{state} of the environment so as to maximize the long-term reward $R_1+\gamma R_2 + \gamma^2 R_3 + ...$. The parameter $\gamma$ is called the \textit{discount factor} and specifies the relative importance of future rewards with regard to the immediate reward.
In general, actions do not only affect the reward that immediately follows, but also the whole chain of reward signals that come after it.
An RL algorithm, then, is a method that uses the agent's interaction with the environment and the obtained rewards to learn better policies over time.
Consequently, RL is a particular type of Machine Learning where the data acquisition is part of the learning process \citep{sutton_reinforcement_1998}.

In RL, the learning agent itself is not necessarily aware of the MDP, or more particularly the transition function. This sets the first distinction between two main approaches of addressing an RL task. They consist of:

$(1)$ Model-based RL approaches use a description of the environment (transition and reward function) that is either known beforehand or learned through experience. In the latter case, Dynamic Programming methods can be used to build an optimal policy. However, having a model of the environment is typically not an option in many use cases, especially if the problem itself changes over time.

$(2)$ Model-free RL approaches do not take into account an explicit model of the environment but rather use trial and error to estimate the state value function $V(s)$ which indicates the expected long-term reward that can be obtained from encountering the state $s$. Alternatively, the action-state value function, $Q(s,a)$ (also called Q-function), has been used in several RL algorithms. It indicates the expected long-term reward that can be obtained by performing action $a$ from state $s$. Values of such a function are called Q-values.

How value functions are updated then constitutes another way of categorizing RL approaches. A very popular approach is  Temporal Difference Learning, where the value of a certain configuration ($V(s)$ or $Q(s,a)$) is updated after encountering the given configuration and then observing the received rewards over a certain time period. The value of the configuration is then adjusted to better match the observed reward sequence and taking into account the expected long-term reward from the end of the sequence. In on-policy methods, the update rule supposes that future actions are taken using the current policy, whereas off-policy methods do not. A well known example of an on-policy temporal difference learning is SARSA (state-action-reward-state-action), for which the update rule is:

\begin{equation}
Q(S_t,A_t) \leftarrow Q(S_t,A_t) + \alpha [R_{t+1} + \gamma Q(S_{s+1},A_{t+1}) - Q(S_t,A_t)].
\label{eq:sarsa}
\end{equation}
In this equation, the ``new'' Q-value of the pair $(S_t, A_t)$ is computed by shifting the ``old'' value towards the most recent observation, which consists of the immediately obtained reward and the estimation of the long-term reward that can be obtained from the new state $S_{t+1}$.
Note that to perform the update, $A_{t+1}$ must be already selected (by the policy). In contrast, the popular off-policy temporal difference learning algorithm, called Q-learning, uses the following update rule: 
\begin{equation}
  Q(S_t,A_t) \leftarrow Q(S_t,A_t) + \alpha [R_{t+1} + \gamma \max_a Q(S_{s+1},a) - Q(S_t,A_t)].  
\label{eq:qlearning}
\end{equation}

It is supposed that the value function is kept in a tabular presentation (the value for each possible input of the value function is stored). The optimal policy can then be derived from the converged Q-values by selecting the action that maximizes the long-term reward from the current state. 

In the learning process, however, it would not be effective to always select the current best action; this would result in the algorithm only exploiting a subset of the potential solutions and not exploring other areas. This is known as the exploration-exploitation dilemma, and various strategies have been used to address it. One of them is the $\epsilon-$greedy method, where at each decision step, a random action is selected with probability $\epsilon$ and the current best action is selected with probability $1-\epsilon$.

Given the possible multidimensional nature of the state values (e.g, a vehicle's state can be comprised of its location, number of passengers, state of charge and current time), in most problems, a tabular representation is difficult to achieve, since the state space can be extremely large. Especially if the state is a vector containing different features, then the size of the state space is exponential to the number of features. This is known as the curse of dimensionality. Another drawback of a tabular representation is that the value of a particular configuration will remain completely unknown until it has been encountered at least once. No generalization is performed from the potential encounters of similar configurations. To solve these issues, value function approximation methods are used. Value functions can be approximated using a more compact representation, for which the number of parameters is considerably less than the number of possible configurations. The approximation then needs to be updated to minimize the error between the observed rewards and the estimated long-term values. Even though each update is typically performed using a given configuration, it updates the values of other configurations. A wide range of value function approximators have been explored in the literature, from linear and polynomial models to Fourier bases and coarse coding. \citet{geist_algorithmic_2013} present a review on the different methods of approximating value functions that are used in RL, along with advantages and drawbacks of each method. 

Recently, a particular kind of approximation has become increasingly popular, that consists of Artificial Neural Networks (ANNs). An ANN, with multiple layers (called deep ANN) can be used to approximate a very wide range of non-linear functions and has proven to be very efficient in supervised learning tasks (in which the learning is performed from labelled data). The interest of using deep ANNs for RL gave rise to the branch of Deep-RL, which is currently extensively used. Q-learning based methods where the Q-function is represented with a deep neural network (deep Q-network) are called deep Q-learning methods.

In contrast to extracting the policy from the learned value function, another approach is to represent the policy as the parameterized function $\pi(a | s, \theta) \in [0,1]$ (indicating the probability of selecting an action given the state and the policy parameters). The parameters $\theta$ of the policy are then updated during the interaction with the environment in an attempt to converge to the optimal policy. RL methods using this approach are known as Policy Gradient Methods, among which we find the generic REINFORCE algorithm \citep{williams_simple_1992} which involves generating multiple sequences of states, actions and rewards (keeping the policy constant during each sequence) and then using the following update rule for $\theta$.

\begin{equation}
    \theta_{t+1} = \theta_t + \alpha G_t \frac{\nabla_{\theta} \pi (A_t | S_t, \theta_t)}{\pi (A_t | S_t, \theta_t)}
\end{equation}

Where $G_t=R_t+\gamma R_{t+1} + \gamma^2 R_{t+2} + ...\ldots$ is the sum of rewards in the following of the sequence. This update rule updates the policy parameters such as to increase the probability for an action at a certain state if it yielded a positive cumulative reward during the sequence and decreases it otherwise. 

Actor-Critic methods \citep{konda_actor-critic_1999} combine a parameterized policy function (actor) with a parameterized value function (critic) $v(s, \text{w})$ that is learned and used to guide the update of the policy. The one-step actor critic algorithm, which does not require sampling a whole episode before the update.It uses $\delta_t = R_{t+1} + \gamma v(S_{t+1}, \text{w}_t) - v(S_t, \text{w}_t)$, which indicates the difference between the observed and expected reward, to update both the value function and the policy function
\begin{equation}
    \theta_{t+1} = \theta_t + \alpha^{\theta} \delta_t \frac{\nabla_{\theta} \pi (A_t | S_t, \theta_t)}{\pi (A_t | S_t, \theta_t)}
\end{equation}

\begin{equation}
        \text{w}_{t+1} = \text{w}_t + \alpha^{\text{w}} \delta_t \nabla_{\text{w}} v(s, \text{w}_t)
\end{equation}

This method presents the advantage of taking into account the experience accumulated by the critic throughout the whole learning process, thus reducing the chances of iterating around a local optimum for too long.

\subsection{Sequential decision-making under uncertainty}
\label{powell}

For MoD applications, several RL approaches have been presented in literature. To discuss them in a systematic way them, we refer to \cite{powell_unified_2019} who proposes a modelling framework for sequential decision-making problems and methods for solving them. The framework aims to unify all the methods and fields that address the general idea of sequential decision-making, such as stochastic optimization, reinforcement learning and optimal control. \cite{powell_unified_2019} suggests that sequential decision-making methods can fit into four classes.
    
In \textit{Policy Function Approximations} (PFA), the policy is a parameterized function of the state.  The goal is then to find the values of the parameters that maximize the objective function. Consider the task of controlling a fleet of electric vehicles. Deciding when to send them to charging stations can be parameterized with a threshold $\theta$ on the state of charge. $\theta$ is the policy parameter that is learned.

In \textit{Cost Function Approximations} (CFA), an embedded optimization problem is solved to find the best decision at each time (e.g., a linear programming algorithm that takes direct observations as input). However, the input of this optimization is parameterized, and good parameter values need to be learned in order to maximize the performance. For example, the expected occupancy of charging stations can be learned by time of day and used to parametrize an optimization algorithm to schedule slots at charging stations. 

Third, \citet{powell_unified_2019} finds that \textit{Value Function Approximation} (VFA) techniques constitute a large share of the literature on sequential decision-making. 
In this area, the goal is to build the value function and then derive the optimal policy. In contrast to CFA methods, what is learned here is the actual interest that situations represent (obtained rewards). These values are then used to orient the actions in order to reach good situations in which the reward is maximized. For instance, a VFA method can learn the expected reward of sending vehicles to charging stations as a function of the current state (time of the day and state of charge).

Finally, in \textit{Direct Lookahead Approximations} (DLA), policies explicitly consider possible future situations, often by using a lookahead model that approximates the true behavior of the problem. Instead of approximating the value function, here what is approximated is the problem itself and the future states that will be encountered. The decision can then be taken by looking ahead using the approximated model. Such an approach can be used to learn a model for the evolution of the state of charge depending on the decisions, which can be taken such that they yield the best outcome according to the model.


%% file: 3_Literature.tex
\section{Literature}



In this literature review, we focus on papers with an emphasis on the use of RL for the operation of MoD systems and use cases that attempt to reflect real-world settings. Our review is not exhaustive as we identified recent and relevant papers for our research. The literature was searched with the keywords \textit{Mobility on Demand}, \textit{taxi}, \textit{ride-hailing}, \textit{operation}, \textit{fleet management}, \textit{rebalancing}, \textit{relocation}, \textit{repositioning}, \textit{dispatch}, \textit{vehicle assignment} in combination with \textit{reinforcement learning} and \textit{learning}. To extend the set of covered papers, we performed a manual snowball search through the referenced articles. 

We review the approaches under three axes: (i) the algorithmic core of the approach, i.e., the RL method used, the task, the settings and the algorithmic fit in the framework presented in Section \ref{powell}; (ii) the evaluation methodology, i.e., exploring how the algorithm is assessed, what parameters are studied (sensitivity analyses) and against what approaches it is compared; and (iii) the use case aspect, determining whether the study considers an urban vs. rural setting, and whether it takes into account congestion and public transport. 

We structure this section with respect to the MoD tasks that are performed by the approaches presented in the papers we reviewed: rebalancing only, dispatch only and joint rebalancing and dispatch.

\subsection{Rebalancing}

In \citep{fluri_learning_2019}, a reinforcement learning technique is used to perform the rebalancing task. The problem is characterized by a set of disjoint zones $z \in Z$ and customers waiting in each zone $C_z$. In both studied approaches, the goal of the algorithm is to decide the target numbers of on-demand vehicles $V_z$ in each zone $z$ by using the negative sum of waiting customers $-\text{card}(C_z)$ as a reward. 

Two variants were implemented, one using a classical tabular reinforcement learning where the parts of the network are all considered together on the same level in the value function. The second variant is a cascaded reinforcement learning, where the network is hierarchically divided using the Lloyd k-means algorithm, with $k=2$ areas on each level \citep{lloyd1982least}. The training part is then performed from the upper level to the lower ones. A Q-Learning approach is used in both approaches, with an epsilon-greedy policy. An integer linear program is then used to compute the rebalancing decisions for each vehicle in order to match the demand in each zone while minimizing the overall travel distance for the vehicles. Consequently, Reinforcement Learning is used in this approach to build the input of an optimization problem. Therefore, we classify this method as a CFA. 

The model is evaluated using publicly available data from the city of San Francisco with the agent based simulation framework AMoDeus \citep{ruch2018amodeus}. The paper does not present sensitivity analyses, but rather focuses on the comparison of the two tested approaches with other control theoretical algorithms that exist in the literature \citep{pavone2012robotic}. The results on the tested cases show the advantage of using the cascaded reinforcement learning algorithm.

\vspace{0.2cm}\citet{wen_rebalancing_2017} used a Deep Q-Learning approach for the rebalancing of on-demand vehicles. The optimal rebalancing problem definition in the paper also features a network divided into a set $z\in Z$ of disjoint areas where numbers of incoming requests are assumed to follow a Poisson process $A_z \sim \text{Poisson}(\lambda_z \Delta T)$, with $\lambda_z$ being the arrival intensity for zone $z$, and $\Delta T$ the frequency of rebalancing. The decision variables consist of the matrix $r_{ij}$ indicating the number of vehicles to be rebalanced from zone $i$ to zone $j$, while $r_{i,} = \sum_{j\in Z} r_{ij}$ is the number of vehicles available in zone $i$. The objective is then to maximize $\sum_{j \in Z} b_{j}(v'_j) - \sum_{i,j \in Z}c_{i,j}r_{i,j}$. Here, $v'_j$ is the expected number of assignable vehicles that will be available in zone $j$ at rebalancing time and $b_{j}(v'_j)=\sum_{k=0} \text{min}(k, v'_j) \Pr(A_j=k)$ is the expected number of requests that can be served if there are $v'_j$ vehicles in the same zone. $c_{i,j}$ indicates the cost of moving a vehicle from zone $i$ to zone $j$, it is set to $cd_{i,j}$ if zone $j$ is reachable from zone $i$ within $\Delta T$ (with $d_{i,j}$ being the distance between the two zones); otherwise, it is set to a large constant $\overline{c}$. 

Under the assumption of deterministic travel times and unchanged vehicle routes, and knowing the rebalancing decision, $v'_j$ is defined as follows: the vehicles that should be at zone $j$ at rebalancing time are considered with different weights according to the likelihood that a seat will be available depending on their current occupancy. The respective weights are $1$, $0.4$, $0.2$, $0.1$, $0$ for loads $0$, $1$, $2$, $3$, $4$ given 4-seated-vehicles. These assumptions result in a Mixed Integer Nonlinear Programming (MINLP) problem that is solved approximately using a combination of incremental-optimal and branch-and-bound methods. This solving approach is not detailed further in the paper and is referred to as Heuristic Optimal Rebalancing (HOR).

In addition to comparing the deep Q-learning approach against HOR, the authors defined a Simple Anticipatory Rebalancing (SAR) strategy that rebalances a vehicle from zone $i$ to a zone $j \in \overline{Z_i}$ where $\overline{Z_i}$ represents the current and neighboring zones of $i$ ($i \in \overline{Z_i}$). The rebalancing zone is then sampled using probability $\Pr(\text{vehicle moves to j}) = \lambda_j / (\sum_{j' \in \overline{Z_i}}  \lambda_{j'})$. 

For the deep Q-learning algorithm, each vehicle considers its neighboring areas as the environment and their idle vehicles, in-service vehicles and predicted demands as constituting the state. The action is then chosen with an $\epsilon-$greedy policy,  amongst one or none of the neighboring areas; if none is chosen, the vehicle does not rebalance. The reward design in this approach is twofold: if the vehicle is assigned, the setting is compared to the one without rebalancing and the saved wait time is taken as the reward (how this is computed is not detailed in the paper); and if the vehicle is not assigned, a constant penalty is applied. We note that this reward does not directly reflect the objective function that is presented in the optimal rebalancing problem formulation. Since the output of the algorithm is the straightforward location to which the vehicle needs to relocate to, we classify this approach as a VFA.

The algorithms are benchmarked on an abstract map with varying sizes under three different scenarios for the demand: uniformly distributed trips origins and destinations; two areas concentrating origins and two areas concentrating destinations; uniformly distributed origins with one fixed destination. The fleet sizes for the three different map sizes are $20$, $125$ and $810$ respectively. For each setting, observed performances for HOR were better than for Deep Reinforcement Learning which were better than SOR. The three algorithms were all better than the absence of rebalancing. In terms of computation times, the increase was more drastic for HOR than for SOR and DQN. The rebalancing methods were then tested in a use case of shared MoD in Orpington, London using travel data spanning a 10 years period. The results in this use case showed similar relative performance between the algorithms. The different tests were performed on the agent-based modeling platform detailed in \citep{wen_transit-oriented_2018}. Due to the assumptions of fixed travel times that were taken, we consider that this work does not take congestion into account. 

\vspace{0.2cm}
In \citep{yoshida_distributed_2021}, a decentralized deep Q-learning is used for the rebalancing task. The network is divided into a grid $Z$ and each vehicle $v$ considers a service area $S_v \subseteq Z$. The goal is then to adapt the service area for each vehicle. This is performed using a DQN for which the input (state) consists of the current demand $d_z$ in each zone $z \in Z$; the demand forecast estimation $d'_z$ in the next time period and the number of vehicles in the same area $n_v = \text{card}\{v' |v' \text{ located in zone z}, z \in S_v\}$ allowing the vehicles to take each other into account. Adapting the service area can be performed with six actions: enlarge, shrink, move up, down, left, right and stay (leaving the service area unchanged). The reward recieved by a vehice $v$ is defined as $r_v = w_{b}b_{v} - w_{d}d_{v} - w_{f}f_{v}$ where $b_v$ indicates the number of travelers assigned to $v$, $d_v$ the distance between $v$ and them and $f_v=1$ if the last action selected by $v$ is stay and $f_v=0$ otherwise. The actions are selected following an $\epsilon$-greedy policy with a linearly decaying $\epsilon$. When idle, a vehicle is simply relocated to the center of its service area. We classify this approach as a VFA. 

The algorithm proposed here is evaluated using the public benchmark dataset of taxi trips in Manhattan in a simulated grid environment. This study shows the advantage of using the proposed approach in comparison to the one presented in \citep{yoshida2020multi}, another RL approach \citep{wen_rebalancing_2017}, and an approach that uses forecast data to move unassigned vehicles to areas with shortages using linear programming \citep{miao2016taxi}.

\vspace{0.2cm}
In contrast to previously mentioned works where the network is discretized into zones, \citet{kim_optimizing_2021}
use a Deep Q-Learning approach in which the network is modelled as a directed graph $G=(V,E)$. $V$ and $E$ respectively denote the set of intersections and the set of roads linking them. At each time $t$ and for each road $j\in E$, $v_{j,t}\in \mathbb{N}$ denotes the number of empty vehicles on road $j$, $n_{j,t} \in \mathbb{N}$ the number of waiting travelers and $p_{j, t}$ the speed on $j$. The deep Q-network takes as input the state of the road network that consists of a vector $s_t$ containing the states $s_{j,t}$ $\forall j \in E$ with $s_{j,t}=(v_{j,t}, n_{j,t}, p_{j,t})$. After an action, each vehicle receives a reward value of $1$ if it was assigned to a request and $0$ otherwise. $Q(s_t, j)$ with $j \in E$ is the expected long term reward earned after moving vehicles to $j$ from its neighboring roads. The update of the Q-values is not performed following the standard Q-learning equation, but rather an expected-SARSA (Equation \ref{eq:sarsa}). Using the Q-values, the policy selects stochastically the next road for each vehicle, such as the probability of moving from $j$ to $k$ is determined with the relative interest of road $k$ among all roads adjacent to $j$. Therefore, we categorize this approach as a VFA. The tests of this algorithm were performed in a custom-built simulator. However, the scalability of the method is not assessed as the size of the network considered in the use case has not been explicitly mentioned.

\vspace{0.2cm}
\citet{gammelli2021graph} present a RL algorithm for rebalancing where the network is divided into zones that are then considered in a graph $G = (V, E)$ where each zone is linked to the nearest ones. Travelling from zone $i$ to $j$ ($i,j \in V$) at time $t$ is has a cost $c_{i,j}^t$, a travelling time $\tau_{i,j}^t$, and a profit $p_{i,j}^t$ if the trip is transporting a passenger. The number of requests to travel from $i$ to $j$ at time $t$ is noted $d_{i,j}^t$ and the number of successfully served ones among them is $x_{i,j}^t$. The RL rebalancing method is centralized and the state variable is composed of: (i) the adjacency matrix of the graph (ii) the numbers $m_{i}^t \in [0, M]$ $ \forall i \in V$ of vehicles with M being the fleet size (iii) the projected availability of vehicles $m_{i}^{t'}$ $\forall i \in V, t'\in [t+1,...t+T]$ with $T$ being the planning horizon (iv) the current demand $d_{i,j}^t \forall (i,j) \in E$ and (v) the estimated future demand $d_{i,j}^{t'} \forall (i,j) \in E, t'\in [t+1,...t+T]$. An action consists in choosing the desired distribution $a_{reb}^t = \{a_{reb,i}^t\}_{i \in V}$ of vehicles across the zones ($\sum_{i\in V} a_{reb,i}^t = 1$). The reward $r_{reb}^t$ that is considered in this approach is operator-centered, as it reflects the profit made by the service from served requests and the cost of moving vehicles $r_{reb}^t = \sum_{i,j \in V} x_{i,j}^t (p_{i,j}^t - c_{i,j}^t) - \sum_{(i,j) \in E} y_{i,j}^t c_{i,j}^t$. The rebalancing decision on the zone level, i.e, choosing the number $y_{i,j}^t$ $\forall (i,j) \in E$ of how many vehicles are relocated from zone $i$ to zone $j$, is then performed using linear programming method minimizing the rebalancing cost. We then classify this approach as a CFA. 

The policy is modelled with a graph neural network that allows to generalize beyond the order with which the zones are considered and make abstract conclusions that can be transferred between use cases. The policy is trained following the actor-critic approach and tested on two use-cases related to the cities of New York and Chengdu (China). The authors compare against equal distribution of vehicles among zones; the cascaded Q-learning approach presented in \citep{fluri_learning_2019}; and using feed-forward or a convolutional neural network instead of a graph neural network and the use of model predictive control methods in order to provide an upper bound of performance. In both cases, the presented approach is able to achieve the best reward performance excluding the model predictive control methods. To our knowledge, this paper is the only one in the literature that explores the potential of transferring a policy to either a completely new network or an extended one. The transferred policies achieve less reward than their counterparts learned on the new use case, but are still better than the other learning based approaches.

\subsection{Dispatch}
In the work presented in \citet{guoyang_qin_optimal_2020}, a RL algorithm is used to aid the vehicle assignment task by selecting which vehicles and requests to consider for the decision-making and which to delay to the next decision epoch. Similarly to other works in the literature, this approach considers a network divided in a set of zones $z \in Z$. The approach is centralized, with the state $s(t)$ at time $t$ being modelled as $s(t)=\{\{N_p(z,t), N_d(z,t), \lambda_p(z,t), \lambda_d(z,t)\} | z\in Z\}$ where $N_p(z,t)$ and $N_d(z,t)$ are the number of requests and idle vehicles in zone $z$ at time $t$. $\lambda_p(z,t)$ and $\lambda_d(z,t)$ are the estimated arrival rate of requests and idle vehicles to zone $z$ from $t$ onward. An action $a \in \{0,1\}^{|Z|}$ is a vector composed of binary values $a_z \in \{0,1\}$ for each $z \in Z$ where $a_z=1$ if the requests and drivers of zone $z$ will be considered in the next dispatch decision and $a_z=0$ otherwise. The reward that is considered here is the induced waiting times for the users. 

A policy gradient method is used in this work. The policy is a neural network that takes as input the state $s(t)$. The output layer of the neural network contains $2^{|Z|}$ elements in $[0,1]$  specifying a probability distribution of actions. This probability distribution is then used to sample the action. The weights of the neural network (and thus the policy) are adjusted following the Actor-Critic and Actor-Critic with Experience Replay methods. Selected requests and drivers are matched by solving a Bipartite Matching Problem. We consider this approach to be a CFA. 

This approach was evaluated in a numerical experiment using a real world one week dataset from Shanghai Qiangsheng Taxi collected in March 2011. It has been compared against fixed matching delays in terms of resulting waiting times for the users and showed better performance. However, in this study, the number of considered zones did not exceed $6$ whereas the size of the action space is exponential in the number of zones. 

\vspace{0.2cm}
\citet{enders2022hybrid} present a deep RL approach to address the dispatch of MoD vehicles. 
The problem considered in this work is characterized by a service area modelled as a graph $G=(V,E)$. Each edge $e \in E$ is associated with a weight vector $w_e=(w_e^{d}, w_e^{t})$ where $w_e^{d}$ denotes the distance of the edge and $w_e^{t}$ the time necessary to traverse it. The approach is centralized and the system state at time $t$ is defined as $s_t=(t, (r_t^i)_{i \in \{1,...,R_t\}}, (k_t^j)_{j \in \{1,...,K\}})$ with $R_t$ being the number of travel requests at time $t$ and $K$ the fleet size. A request $r$ is defined as $r=(\omega, o, d)$ with $\omega \in \mathbb{N}_{0} \cup \{\emptyset\}$ is the current waiting time (it is set to $\emptyset$ at pickup), $o \in V$ and $d \in V \backslash \{o\}$ refer to the origin and destination of the request. A vehicle state $k_t^j=(v_t^j, \tau_{t}^{j},  r_{t}^{1,j}, r_{t}^{2,j})$ consists of a position $v_t^j\in V$, the time $\tau_t^j$ left to reach $v_t^j$ (the vehicle can be travelling) and up to  two assigned requests $r_{t}^{1,j}$ and $r_{t}^{2,j}$. Possible actions to perform at time $t$ are tuples $(a_t^1,...,a_t^{R_t})$ where $a_t^i \in \{0,...,K\}$ such as $a_t^i = 0$ if request $r_t^i$ is rejected and $a_t^i = j$ if it is assigned to the vehicle $j$. Only actions that satisfy $a_t^i=j \in \{1,...,K\} \Rightarrow r_{t}^{2,j} = \emptyset$ $\forall i \in \{1,...,R_t\}$ (no request is assigned to a vehicle that already has two requests assigned to it) and $\sum_{i=1}^{R_t} \mathbbm{1}(a_t^i=j) \leq 1$ $\forall j \in \{1,...,K\}$ (at most one new request is assigned to each vehicle). The goal is to maximize the profit generated by the fleet: when picking up a passenger (related to a request $r_t^i=(\omega, o,  d)$), the system generates a revenue $rev(r_t^i) > 0$ if $\omega < \omega_{max}$ and $rev(r_t^i)=0$ otherwise; after moving from node $v$ to node $v'$ through edge $e=(v, v')$, a vehicle generates a cost equal to $w_e^t$. At each time $t$ and after performing the action, the system receives as a reward the sum of revenues generated by picked up requests from which are subtracted the costs generated by moving vehicles.

In contrast to most decentralized RL approaches, where each vehicle is considered as an RL agent, here each vehicle-request pair is considered as a separate RL agent that provides the probability of the request being assigned to the vehicle. This allows to build a weighted bipartite matching graph that is solved to compute the effective assignments. This method is consequently classified as a CFA. The learning is performed following an Actor-Critic method using neural networks for both actor and critic functions. The method is tested using experiments based on the New York Taxi data \citep{tlc2020nyc} and compared against a greedy method and a model predictive control method. The proposed RL method shows better performance in general and more stability across the testing period. 

\subsection{Rebalancing and dispatch}

In \citet{gueriau_shared_2020}, a decentralized Q-learning algorithm for dispatch and rebalancing is studied. The network is divided to a set of zones $z \in Z$. At time $t$, the state of a vehicle $v$ is $s_{v,t} = (l_{v, t}, d_{v}, d'_{v})$ where $l_{v, t} \in \{\text{empty}, \text{partial}, \text{full}\}$ denotes the vehicle's load and $d_{v}$ and $d'_{v}$ are equal to $1$ if there is a pending request in the same zone as $v$ or one of the neighboring zones, respectively. The possible actions for each vehicle to decide from are:
picking up a passenger by choosing the closest open request;
rebalance by choosing from 4 strategies for selecting which neighboring zone to relocate to (the one with most requests, the one with the biggest gap between vehicle number and vehicle demand, the one with most historical requests, the one with the biggest historical gap between vehicle number and vehicle demand); or do nothing.
This allows partially occupied vehicles to rebalance and serve other requests on their route. A vehicle receives a positive reward when picking passengers and no reward otherwise. The decisions can be carried out straightforwardly without further optimization, we consequently consider this approach as a VFA.

The proposed model was evaluated using the SUMO simulator \citep{lopez2018microscopic} on generated ride-requests for Manhattan using the New York City taxi data set over 50 consecutive Tuesdays from July 2015 to June 2016.
The study focused on the morning rush hour.
To take congestion into account, private vehicle trips have been generated using a uniform distribution.
No analysis of the impact of the RL parameters is described in this study.
The performance of the system and its impact were evaluated on three aspects:
The system perspective, with the amounts of served and timed-out requests (requests expire within 10 minutes);
The rider perspective: waiting time, detour time, total travel time;
The vehicle perspective: total Vehicle Miles Traveled (VMT), empty VMT, engaged VMT, shared VMT and also vehicle occupancy.
The algorithm has been compared to simple strategies involving centralized and decentralized vehicle rebalancing (each selecting the request with the longest waiting time) and rebalancing vehicles in the centroid of their predetermined ``home'' zone and ride-sharing.

\vspace{0.2cm}
\citet{al-abbasi_deeppool_2019} present DeepPool, a decentralized Deep Reinforcement Learning approach to learn good dispatch and ride-sharing behaviors for the vehicles on the zones of the network.
The authors first present a detailed mathematical presentation of the problem where the network is divided into zones as is done in other approaches and where vehicles with a least one empty seat can decide to take in new requests. Each vehicle in the system has a state $S_{t,v}$ that consists of the vehicle's location, number of vacant seats, the passenger pickup time and the destination of each passenger.  
The global system state $s_t$ variable comprises the vehicles' states, the estimations of the number of vehicles in each zone in $T$ time steps $V_{t:t+T}$ and the estimation of the future demand in each zone $D_{t:t+T}$. 
Three neural networks are used in this framework:
one to estimate travel times across the network which, when combined with the current vehicles' plans, allows estimating $V_{t:t+T}$;
one to estimate future demands $D_{t:t+T}$;
and lastly a deep Q-network that is fed with the state variables (including the outputs of the other networks) and estimates the Q-values of doing each of the possible actions: (i) where to dispatch each vehicle and, (ii) whether it takes new requests. The reward signal is a weighted sum combining:
$(1)$ the difference between demand and supply;
$(2)$ the total dispatch time of the vehicles;
$(3)$ the overhead caused to passenger travel times by ride-sharing;
$(4)$ the number of used vehicles.

The learning process uses an $\epsilon-$greedy method with a linearly decreasing $\epsilon$ and also a linearly decreasing learning rate $\alpha$. We classify this approach as a VFA. The decisions operated by this algorithm are directly performed by the vehicles without going through another optimization process. In this study, the algorithm is evaluated on the public dataset of taxi trips in Manhattan, New York \citep{tlc2020nyc}.
These data are used to build requests that are fed to a built-in simulator and handled by the service operated by DeepPool.
The main metrics used to assess the performance of the system are the ones present in the objective function detailed above, plus waiting time and the rate of rejected requests.
The estimation of travel times uses historical trip data, which means that traffic and congestion are considered in this study, but the impact of the MoD service on traffic is not.

\vspace{0.2cm}
In \citet{haliem_adapool_2022}, the authors extend the research conducted in \citet{al-abbasi_deeppool_2019} to address the issue of catastrophic forgetting observed in neural networks \citep{kemker2018measuring}. This is performed by considering the environment as changing between a set of models (more specifically transition functions) and employing a change point detection algorithm to switch between models\citep{kj2022change}. Each model is associated with a deep Q-network that is trained with experiences observed in the given context. Consequently, this algorithm learns different policies as well as the appropriate time to use each policy. This approach is used to learn diurnal variations of the demand in a scenario based on a real public dataset of taxi trips in Manhattan, New York City. The results show the advantage of this approach over having only one model.

\vspace{0.2cm}
\citet{mao_dispatch_2020} use an RL approach for combined vehicle assignment and rebalancing. In this work, the service area is divided into a set of zones $Z$. The state variable is a vector consisting of the time of the day $t$; the matrix $R_{i,j}$ with $i,j \in Z$ of number of requests with origin in $i$ and destination in $j$ and a vector $V_i$ indicating the number of vehicles in each zone. The action learned in this approach is a matrix $M_{i,j}$ specifying the number of vehicles to relocate from $i$ to $j$ with $\sum_{j \in Z}  M_{i,j} = V_i$. The vehicle assignment is directly dependent on the relocations, as requests can be served by vehicles with the same origin and destination if available. Otherwise, they will remain pending. Consequently, vehicles can either serve a request if they are assigned to it or simply relocate in the network. The reward signal used in this approach combines unassigned vehicles' relocation costs and users' waiting times.

To avoid the intractability of a large state-action space, a policy-based actor critic method is used, where the policy is a function of the state. In this approach, the policy and critic function are represented with feed forward dense neural networks that are adapted to enforce that the number of dispatched vehicles is not negative, and the constraint on the number of originating vehicles $V_i$ is fulfilled. We therefore put this approach in the PFA category. Two settings were considered for the objective. In the first one, all users' waiting times are considered equally, while in the second one features ``impatient'' users with waiting times that exceed a determined threshold. Those are more weighted higher in comparison to ``patient'' users. The experiments were performed on a dataset of taxi trips in Manhattan. Compared against the REINFORCE algorithm with similar settings and an optimal solution method providing an upper bound, this approach showed to have better results.

\vspace{0.2cm}
\cite{tang_deep_2019} and \citet{tang_value_2021} propose an approach that embeds both vehicle dispatching and rebalancing in one framework while combining both historical data and data acquired online. The algorithm is decentralized among the vehicles and the state $s$ is modelled as $s=(l, u, v)$ where $l$ refers to the vehicle's location, $u$ denotes the real world time stamp and $v$ is a vector comprising other dynamic information (e.g., current supply and demand) and static information (e.g., day of the week, holiday indicator). The vehicles learn to choose an action $a\in \{d, r\}$ (dispatch and rebalancing). The goal is to maximize the discounted sum of rewards $\sum_{j=1}^{T} \gamma^{j-1} r_j$ with $r_j$ denoting the revenue generated by the trip assigned to the vehicle if $a_j=d$ and $0$ otherwise. The dispatching is performed across all selected vehicles in order to maximize the total sum of expected rewards. The rebalancing destination is chosen stochastically for each selected vehicle based on weights derived from the expected value of each alternative.

A key interest of this approach lies in using one single value function for both the dispatch and rebalancing decisions for all the vehicles. The historical data are used to build an offline value function estimation, which is then taken into account alongside online observations to build the usable value function, which is unique for all the vehicles and updated by all their observations. We consider this method to be a CFA, since the dispatch decision is performed by solving an optimization problem that is built using the value function. 

The experiments have been performed in simulation environments built from DiDi's real-world ride-hailing data regarding three different cities over two week-days and two weekend days.   
The \textit{dispatching} performance of the proposed approach was compared to a baseline myopic method, a greedy method, the one presented in \cite{tang_deep_2019}, and the one that received the first prize from the same task in the KDD Cup 2020 RL track competition. Results show that the approach outperformed the others. On the \textit{rebalancing} side, this approach was compared against the winning method in the rebalancing task of the same competition, a human expert policy extracted from historical data, and a deterministic and greedy version of the proposed approach where the location with the highest value is selected. The performance in this task was studied under different fleet sizes, and it is shown that the proposed method outperforms the others consistently in both versions. The approach was also tested on settings that attempt to reflect temporary and considerable changes on the supply and the demand, like the arrival of new vehicles or an event triggering many unexpected travel requests. The results show that the framework is able to adapt well to such situations.

This work is extended in \citep{eshkevari_reinforcement_2022} where results are presented regarding the implementation of the approach in a real-life ride-hailing service operated by DiDi. A/B testings were first conducted in five major cities where drivers switched between the RL algorithm and the baseline methods for periods of three hours and which demonstrated the relative performance of RL. The algorithm was then adopted to be used in one large unspecified city in China and has been in service since late December 2021.

\vspace{0.2cm}
In \cite{liang_mobility-aware_2021}, a Deep Reinforcement Learning approach is used to make combined decisions regarding dispatch, rebalancing and recharging. Here the network is divided into a set of hexagonal zones $z\in Z$. At each time $t$, the sets $R_{z,t}$  of open requests and $I_{z,t}$ of available vehicles in zone $z$ are considered and for each $v \in I_{z,t}$, $soc_{v,t}$ denotes the state of charge of vehicle $v$ at time $t$. The learning takes place on the zone level and the state of each zone $z$ at time $t$ is $s_{z,t}= (t, soc_{v_1,t}, soc_{v_2,t}\ldots)$ with $I_{z,t}=\{v_1, v_2,\ldots\}$. The zone level action consists of the joint actions of the vehicles $v \in I_{z,t}$. A vehicle's action is either to pick up a request, relocate to an adjacent zone, or recharge at a specific station: $a_{v,t} \in R_{z,t} \cup A_z \cup C_z$. $A_z \subset Z$ denotes the zones adjacent to $z$ and $C_z$ denotes the set of charging stations in $z$. The reward $r_{a_{v,t}}$ obtained by a vehicle $v$ after performing an action $a_{v,t}$ is defined as follows. If $a_{v,t} \in R_{z,t}$ then $r_{a_{v,t}} = \sum_{\tau = t}^{t+\Delta t_{a_{v,t}}-1} \gamma^{\tau-1} \frac{P_{a_{v,t}}}{\Delta t_{a_{v,t}}}$ where $\Delta t_{a_{v,t}}$ denotes the duration of the selected request's trip and $P_{a_{v,t}}$ the revenue generated by it. $\gamma$ is a discount factor parameter. If $a_{v,t} \in A_z$ then $r_{a_{v,t}}=0$. And if $a_{v,t} \in C_z$ then $r_{a_{v,t}}=\sum_{\tau = t}^{t+\Delta t_{a_{v,t}}-1} \gamma^{\tau-1} P_{g,\tau} \Delta (soc_{v, \tau} - soc_{v, \tau-1})$ where $P_{z,\tau}$ refers to the price of electricity effective in zone $z$ at time $\tau$.

The joint decisions of the vehicles in each grid then feed into a binary linear programming algorithm that determines the optimal manner to implement the decisions (making this approach a CFA) while respecting constraints of at most one vehicle per request and one vehicle per charging station. The overall objective is to maximize the revenue of the service (and minimize cost). This algorithm is compared against solving the same binary linear programming problem but with actions determined by heuristics for dispatch, recharging and rebalancing. An interesting feature of this study is that the response of the algorithm to various electricity pricing schemes was studied (fixed prices, prices depending on time and/or location).

\vspace{0.2cm}
\citet{castagna_demand-responsive_2021} use a decentralized RL algorithm for rebalancing and dispatch of a ride-shared MoD system. The algorithm is decentralized and a vehicle state $s_v=(l_v, d_v, e_v, p_v)$ consists of its location $l$ (latitude and longitude), its next destination $d$ (the closest destination for on-board passengers) and the number of empty seats $e$. Additionally, $p_v= (s_r^1, s_r^2, s_r^3)$ is the vehicle's perception vector that includes the information related to the $3$ closest requests. $s_r^i=(l_r^i, d_r^i, n_r^i)$ with $l_r^i$, $d_r^i$ and $n_r^i$, respectively, referring to the request's pickup location, its drop-off location and number of passengers. The action space consists of five alternatives: rebalance, picking up one of three possible requests, and dropping off the passenger(s) with the closest destination. The reward is set to favor ride-sharing by giving a higher reward for picking-up a passenger when the vehicle is already occupied than when it is not. Impossible actions (drop-off when the vehicle is empty or pick-up when the perception vector is empty) are penalized. Like other approaches, the rebalancing procedure considers a set of discrete zones dividing the network. However, here, the set of zones is not fixed beforehand and is computed at each rebalancing decision using an Expectation-Maximization technique based on pending travel requests. Consequently, open requests are dynamically clustered into spatial zones. These clusters are then considered for rebalancing. Each vehicle samples a target zone for rebalancing with a probability equal to the ratio of requests contained in the zone.  

In this approach, a good policy is learned by using a proximal policy optimization method \citep{schulman2017proximal}, making it a PFA method. The approach is tested with the New York Taxi dataset under various configurations characterized by enabling or disabling ride-sharing and by the rebalancing method that is used (no rebalancing, rebalancing with fixed zones, rebalancing with dynamically computed zones). The results show the advantage of using both ride-sharing and dynamic zones computation for enhancing number of served requests and lowering users' wait times.

\input{table_4.tex}

%% file: table_4.tex
\begin{landscape}
\begin{table}[]
\centering
\resizebox{\linewidth}{!}{\begin{tabular}{|c|clcc|c|cccc|cc|}
		\hline
		\textbf{Reference}                              & \multicolumn{4}{c|}{\textbf{Algorithmic aspects}}                                                                         &                                                       & \multicolumn{4}{c|}{\textbf{Use case}}                                                                                                                                          & \multicolumn{2}{c|}{\textbf{Evaluation}}                                                                                                                              \\ \hline
		& \multicolumn{1}{c|}{\textbf{Rebalancing}} & \multicolumn{1}{l|}{\textbf{Dispatch}} & \multicolumn{1}{c|}{\textbf{Ridecharing}} & \textbf{Recharging} & \begin{tabular}[c]{@{}c@{}}\textbf{Policy}\\ \textbf{type}\end{tabular} & \multicolumn{1}{c|}{\textbf{Congestion}} & \multicolumn{1}{c|}{\textbf{PT}} & \multicolumn{1}{c|}{\textbf{Environment}} & \begin{tabular}[c]{@{}c@{}}\textbf{Study} \\ \textbf{area}\end{tabular}                 & \multicolumn{1}{c|}{\textbf{Simulation}}                                                             & \begin{tabular}[c]{@{}c@{}}\textbf{Sensitivity}\\ \textbf{analyses}\end{tabular} \\ \hline
		\cite{fluri_learning_2019}             & \multicolumn{1}{c|}{\checkmark}         & \multicolumn{1}{l|}{}       & \multicolumn{1}{c|}{}          &          & CFA                                                   & \multicolumn{1}{c|}{}         & \multicolumn{1}{c|}{} & \multicolumn{1}{c|}{urban}       & \begin{tabular}[c]{@{}c@{}}San \\ Francisco\end{tabular}                & \multicolumn{1}{c|}{\begin{tabular}[c]{@{}c@{}}MATSim \\ (AMoDeus)\end{tabular}}            & AP                                                             \\ \hline
		\citet{wen_rebalancing_2017}           & \multicolumn{1}{c|}{\checkmark}         & \multicolumn{1}{l|}{}       & \multicolumn{1}{c|}{\checkmark}         &          & VFA                                                   & \multicolumn{1}{c|}{}         & \multicolumn{1}{c|}{} & \multicolumn{1}{c|}{urban}       & \begin{tabular}[c]{@{}c@{}}Orpington,\\ London\end{tabular}             & \multicolumn{1}{c|}{\begin{tabular}[c]{@{}c@{}}grid \\ environment\end{tabular}}            & SP                                                             \\ \hline
		\cite{yoshida_distributed_2021}        & \multicolumn{1}{c|}{\checkmark}         & \multicolumn{1}{l|}{}       & \multicolumn{1}{c|}{\checkmark}         &          & VFA                                                   & \multicolumn{1}{c|}{}         & \multicolumn{1}{c|}{} & \multicolumn{1}{c|}{urban}       & Manhattan                                                               & \multicolumn{1}{c|}{\begin{tabular}[c]{@{}c@{}}grid \\ environment\end{tabular}}            & AP$+$SP                                                        \\ \hline
		\cite{kim_optimizing_2021}             & \multicolumn{1}{c|}{\checkmark}         & \multicolumn{1}{l|}{}       & \multicolumn{1}{c|}{}          &          & VFA                                                   & \multicolumn{1}{c|}{\checkmark}        & \multicolumn{1}{c|}{} & \multicolumn{1}{c|}{urban}       & Seoul                                                                   & \multicolumn{1}{c|}{\begin{tabular}[c]{@{}c@{}}built-in \\ simulator\end{tabular}}          &                                                                \\ \hline
		\cite{gammelli2021graph}               & \multicolumn{1}{c|}{\checkmark}         & \multicolumn{1}{l|}{}       & \multicolumn{1}{c|}{}          &          & CFA                                                   & \multicolumn{1}{c|}{}         & \multicolumn{1}{c|}{} & \multicolumn{1}{c|}{urban}       & \begin{tabular}[c]{@{}c@{}}New York \\ and \\ Chengdu\end{tabular}      & \multicolumn{1}{c|}{\begin{tabular}[c]{@{}c@{}}built-in \\ simulator\end{tabular}}          & MP                                                             \\ \hline
		\cite{guoyang_qin_optimal_2020}        & \multicolumn{1}{c|}{}          & \multicolumn{1}{l|}{\checkmark}      & \multicolumn{1}{c|}{}          &          & CFA                                                   & \multicolumn{1}{c|}{}         & \multicolumn{1}{c|}{} & \multicolumn{1}{c|}{urban}       & Shanghai                                                                & \multicolumn{1}{c|}{\begin{tabular}[c]{@{}c@{}}built-in \\ \\ simulator\end{tabular}}       & AP                                                             \\ \hline
		\cite{enders2022hybrid}                & \multicolumn{1}{c|}{}          & \multicolumn{1}{l|}{\checkmark}      & \multicolumn{1}{c|}{\checkmark}         &          & CFA                                                   & \multicolumn{1}{c|}{}         & \multicolumn{1}{c|}{} & \multicolumn{1}{c|}{urban}       & New York                                                                & \multicolumn{1}{c|}{\begin{tabular}[c]{@{}c@{}}grid \\ environment\end{tabular}}            & AP$+$SP                                                        \\ \hline
		\cite{gueriau_shared_2020}             & \multicolumn{1}{c|}{\checkmark}         & \multicolumn{1}{l|}{\checkmark}      & \multicolumn{1}{c|}{\checkmark}         &          & VFA                                                   & \multicolumn{1}{c|}{\checkmark}        & \multicolumn{1}{c|}{} & \multicolumn{1}{c|}{urban}       & Manhattan                                                               & \multicolumn{1}{c|}{Sumo}                                                                   & SP                                                             \\ \hline
		\cite{al-abbasi_deeppool_2019}         & \multicolumn{1}{c|}{\checkmark}         & \multicolumn{1}{l|}{\checkmark}      & \multicolumn{1}{c|}{\checkmark}         &          & VFA                                                   & \multicolumn{1}{c|}{}         & \multicolumn{1}{c|}{} & \multicolumn{1}{c|}{urban}       & Manhattan                                                               & \multicolumn{1}{c|}{\begin{tabular}[c]{@{}c@{}}built-in \\ simulator\end{tabular}}          & SP                                                             \\ \hline
		\cite{haliem_adapool_2022}             & \multicolumn{1}{c|}{\checkmark}         & \multicolumn{1}{l|}{\checkmark}      & \multicolumn{1}{c|}{\checkmark}         &          & VFA                                                   & \multicolumn{1}{c|}{}         & \multicolumn{1}{c|}{} & \multicolumn{1}{c|}{urban}       & Manhattan                                                               & \multicolumn{1}{c|}{\begin{tabular}[c]{@{}c@{}}built-in \\ simulator\end{tabular}}          & AP                                                             \\ \hline
		\citet{mao_dispatch_2020}              & \multicolumn{1}{c|}{\checkmark}         & \multicolumn{1}{l|}{\checkmark}      & \multicolumn{1}{c|}{\checkmark}         &          & PFA                                                   & \multicolumn{1}{c|}{\checkmark}        & \multicolumn{1}{c|}{} & \multicolumn{1}{c|}{urban}       & Manhattan                                                               & \multicolumn{1}{c|}{\begin{tabular}[c]{@{}c@{}}grid \\ environment\end{tabular}}            &                                                                \\ \hline
		\cite{tang_value_2021}                 & \multicolumn{1}{c|}{\checkmark}         & \multicolumn{1}{l|}{\checkmark}      & \multicolumn{1}{c|}{}          &          & CFA                                                   & \multicolumn{1}{c|}{\checkmark}        & \multicolumn{1}{c|}{} & \multicolumn{1}{c|}{urban}       & \begin{tabular}[c]{@{}c@{}}undefined \\ cities \\ in China\end{tabular} & \multicolumn{1}{c|}{\begin{tabular}[c]{@{}c@{}}DiDi \\ simulation \\ platform\end{tabular}} & MP                                                             \\ \hline
		\cite{liang_mobility-aware_2021}       & \multicolumn{1}{c|}{\checkmark}         & \multicolumn{1}{l|}{\checkmark}      & \multicolumn{1}{c|}{}          & \checkmark        & CFA                                                   & \multicolumn{1}{c|}{\checkmark}        & \multicolumn{1}{c|}{} & \multicolumn{1}{c|}{urban}       & \begin{tabular}[c]{@{}c@{}}Haikou,\\ China\end{tabular}                 & \multicolumn{1}{c|}{\begin{tabular}[c]{@{}c@{}}built-in \\ simulator\end{tabular}}          & AP$+$SP                                                        \\ \hline
		\cite{castagna_demand-responsive_2021} & \multicolumn{1}{c|}{\checkmark}         & \multicolumn{1}{l|}{\checkmark}      & \multicolumn{1}{c|}{\checkmark}         &          & PFA                                                   & \multicolumn{1}{c|}{\checkmark}        & \multicolumn{1}{c|}{} & \multicolumn{1}{c|}{urban}       & Manhattan                                                               & \multicolumn{1}{c|}{\begin{tabular}[c]{@{}c@{}}built-in \\ simulator\end{tabular}}          &                                                                \\ \hline
	\end{tabular}
}
\caption{Overview of the detailed papers and the features of presented approaches. Policy types: see section \ref{powell}. Sensitivity analyses: Approach Parameters (AP), Scenario Parameters (SP), Multiple Scenarios (MP) }
\label{table}
\end{table}
\end{landscape}

%% file: 4_Analysis.tex
\section{Analysis}

In this section, we present our analysis of the literature with regard to our points of interest: The algorithmic basis; the use case aspects; and the methodology that is followed to study the algorithm's performance. In total, 14 papers have been analyzed based on the previous section. Their key characteristics are summarized in Table \ref{table}.

\subsection{Algorithmic aspects}

We first consider the MoD operation sub-tasks that are addressed in the literature. Empty vehicle rebalancing is the operation task that is studied the most. It is studied exclusively in five of the detailed papers and combined with vehicle assignment in seven other papers. The dispatching task is studied in isolation only in \citep{enders2022hybrid} and \citep{guoyang_qin_optimal_2020}.

Regarding service characteristics, ride-sharing is taken into account in eight papers, whereas the necessity of recharging vehicles is considered only in \citep{liang_mobility-aware_2021}.

Going further into detail, we notice that even for approaches that address the same MoD operation sub-task with similar features, the lower level definitions of the problem can differ and service assumptions can vary strongly. For instance, \cite{wen_rebalancing_2017} and \cite{yoshida_distributed_2021} both study the relocation task with ride-sharing. However, the former considers the exact areas that the vehicles will be relocated to while the latter operates on the vehicle's service area that is shrunk, enlarged or moved. Moreover, even when the addressed tasks are the same, the rewards that are maximized can vary widely. For instance, \citet{fluri_learning_2019} uses a reward that reflects the number of waiting users (encouraging the service to serve all requests) while \citet{gammelli2021graph} considers operator profit (offsetting revenue and cost) as a reward. 

Regarding the framework for sequential decision-making presented in section \ref{background}, VFA and CFA approaches are used in six of the detailed papers, respectively. PFA techniques are underrepresented, with only two papers \citep{mao_dispatch_2020, castagna_demand-responsive_2021}. However, the analyzed approaches show the potential of such methods and, hence, point towards promising future research. In terms of representing the value function, only two papers use a classical tabular representation \citep{fluri_learning_2019, gueriau_shared_2020}. All other papers make use of (Deep) Neural Network approximators.

\subsection{Use cases}

All examined papers consider an urban environment with highly dense cities such as San Francisco \citep{fluri_learning_2019}, London \citep{wen_rebalancing_2017} and Manhattan (six out of fourteen). The potential of MoD in rural setting has been demonstrated in the literature \citep{sieber2020improved} and its operation on such use cases using RL techniques is worth investigating in future research. A challenge thereby is to obtain relevant benchmarking data, which could be obtained from individual mobility traces or synthetic data \citep{horl_synthetic_2021} based on dynamic demand simulations that integrate traveler behavior \citep{horl_simulation_2021}.

Six studies consider the presence of private vehicles and congestion. The presence of public transports is never considered throughout the analyzed literature. Consistently integrating the surrounding transport system, hence, still poses a challenge in RL-based approaches and constitutes a pathway for future research. Furthermore, apart from differentiating between delayed and non-delayed passengers \citep{mao_dispatch_2020}, only homogeneous vehicle fleets and passengers are considered in the analyzed papers. However, heterogeneous dispatching, which is relatively rare \citep{molenbruch_typology_2017}, might benefit from the flexible problem formulations in RL.

\subsection{Evaluation methodology}

To simulate vehicle movements, some papers use custom-built simulators.\citet{guoyang_qin_optimal_2020} and \citet{al-abbasi_deeppool_2019} make use of a detailed and realistic road network representation, whereas \citet{yoshida_distributed_2021} supposes a simple grid network. Two of the analyzed papers use  open-source agent-based simulators (MATSim and SUMO) which arguably increase the reproducibility of the presented approaches and allow the integration and coupling of the on-demand systems with the rest of the transport system. 

Most of the sensitivity analyses are performed by varying the parameters of the presented algorithms (six out of fourteen). These analyses regard the parameters that are at the core of the algorithms, whereas environmental parameters, in particular regarding the discretization of the study areas and networks, are not considered. However, they could give useful insights on the performance of the approach and its scalability. Regarding use case parameters, most of the analyses concern the MoD fleet size (five of fourteen).

While some of the presented approaches use abstract problem formations that, hypothetically, can be transferred from one use case to another, only one example \citep{gammelli2021graph} is presented where transferability is demonstrated and quantified. Given that transferability is frequently cited as a potential major advantage of RL-based methods, it is surprising to not see analyses of the excess learning effort (transfer costs) from one case to another more often. We, hence, encourage further research in that direction. 

In terms of benchmarking the performance of the approaches, we observe that new approaches are still rarely tested against other RL-based methods and that authors fall back to comparison with classic methods. This is partly a result of lacking openness in the approaches as only few of them are easily reproducible by other researchers for comparison. Among the papers reviewed in this work, only \citet{gammelli2021graph} provides access to the code allowing to reproduce the results through an open repository. The rarity of comparisons between RL approaches can also be explained by a lack of comprehensive benchmarks in the literature and the early stage of RL-based fleet control methods, but should be considered in the future. The only structured benchmark of RL based algorithms for MoD operation that we have encountered in our review was the KDD RL cup competition, to which \cite{tang_value_2021} has been submitted.

\section{Conclusion}

To summarize, we state that literature on RL-based fleet operations is rapidly emerging, with many approaches having been published in the past three years as of early 2023. 

Given the emerging character of the research field, the approaches that have been analyzed in the present paper cover a large range of ideas and methods with strongly varying problem formulations tackling the problems of vehicle dispatch, rebalancing and recharging. The respective approaches have been summarized and systematized in the present review.

For future research, we identify the need for more consistent benchmarking of novel approaches against other learning-based techniques and integration of the developed control algorithms into open-source platforms for increased comparability. Furthermore, given the high adaptability and flexibility of RL-based approaches, we see strong potential in widening the research towards use cases in heterogeneous and rural contexts. A promising, but currently rarely seen, topic is the potential of transferability in the context of MoD systems.





%% file: Acknowledgements.tex
\section*{Acknowledgements}

This work has been supported by the French government under the ``France 2030'' program, as part of the
Anthropolis research project at the SystemX Technological Research Institute.

\section*{Author contributions (CRediT)}

Tarek Chouaki: Conceptualization; Methodology; Investigation; Writing - Original Draft 

Sebastian Hörl: Conceptualization; Methodology; Investigation; Writing - Review and Editing; Supervision

Jakob Puchinger: Conceptualization; Methodology; Investigation; Writing - Review and Editing; Supervision; Project administration